\documentclass[preprint,review,12pt,sort&compress,a4paper]{elsarticle}

\usepackage{graphicx}
\usepackage{epsfig}
\usepackage{multirow}
\usepackage{amssymb}
 \usepackage{lineno}
\journal{Journal of Nuclear Materials}

\begin{document}

\begin{frontmatter}


\title{Atomistic Mechanism from Vacancy Trapped H/He Atoms to Initiation of Bubble in W under Low Energy Ions Irradiation}


\author{Yu-Wei You$^{a}$, Xiang-Shan Kong$^{a}$, Q. F. Fang$^{a}$, Jun-Ling Chen$^{b}$, G.-N. Luo$^{b}$, C. S. Liu$^{a,\ast}$, B. C. Pan$^{c,\dagger}$, and Y. Dai$^{d,\ddag}$\footnotetext{$\ast$csliu@issp.ac.cn, $\dagger$bcpan@ustc.edu.cn, $\ddag$yongdai@psi.ch}}


\address{$^{a}$Key Laboratory of Materials Physics, Institute of Solid
State Physics, Chinese Academy of Sciences, P. O. Box 1129, Hefei
230031, P. R. China

$^{b}$Institute of Plasma Physics, Chinese Academy of Sciences,
Hefei 230031, P. R. China

$^{c}$Hefei National Laboratory for Physical Sciences at Microscale and Department of Physics, University of Science and Technology of China, Hefei 230026, P. R. China

$^{d}$Spallation Neutron Source Division, Paul Scherrer Institut, 5252 Villigen PSI, Switzerland}

\begin{abstract}
With the first-principles calculations of H and He induced energetics change we demonstrate that in W the accumulation of H (up to 9) and He (up to 4) in a single vacancy (V) surprisingly reduce the formation energy of first and second nearest vacancy (as low as $\sim$0 eV), which gives the direct evidence of V-H(He) complex mutation mechanism from V-H$_{n}$(He$_{n}$) to V$_{2}$-H$_{n}$(He$_{n}$) and with the potential to lead to the growth of H/He-vacancy complexes: an initial step to H and He bubble. This finding well explains the long-standing problem of why H and He bubbles being produced on W surface exposed to low-energy (far lower than displacement threshold energy) D or He ions irradiation. The further identified repulsive (attractive) interaction between V-H$_{12}$(V-He$_{14}$) and additional H(He) illustrates the experimentally observed big difference of deposition depth of H (${\mu}m$) and He ($\sim$100{\AA}) bubbles in W even the migration rate of He is far larger than that of H.
\end{abstract}


\end{frontmatter}


\section{Introduction}
Tungsten (W), a high-Z material, has been chosen to be the plasma facing material (PFM) in the next step fusion device ITER \cite{Iter} due to its excellent properties of high melting point and low sputtering yield. However, as a PFM, W must be exposed to extremely high fluxes of deuterium (D), tritium and helium (He) ions and neutrons, which directly leads to the displacement damage, bubble formation, and ultimate failure of the material \cite{Gilliam1, Gilliam2, Ueda, Xu}. Therefore it is crucial to understand hydrogen (H)/He-metal-atoms and H/He-defects interactions in the material.

Generally, defects in materials such as dislocations, grain boundaries and vacancies can act as traps for H and He, and origins of H and He bubbles. Among these defects, vacancy receives much more attention. A series of works by Fukai \emph{et al}. indicated that H can stabilize and increase the concentration of vacancy because the vacancy formation energies in metals are reduced substantially due to the insertion of H atoms \cite{Fukai1, Fukai2, Fukai3, Fukai4}. The role of vacancy on trapping H to form the V-H$_{n}$ complex and the maximum number of $n$ in the complex were emphasized in the cases of many metals such as Pd \cite{Fukai4, Nordlander, Vekilova}, Al \cite{Lu, Wolverton, Ismer}, Fe \cite{Nordlander, Nazarov, Tateyama} and W \cite{Liu, Jiang, Ohsawa, Heinola}. Nucleation free energies evaluated with density function theory indicated that H trapping assists the divacancy formation in bcc W crystal \cite{Kato}. No doubt, these contribute enormously to our understanding of H bubble formation and blistering, but the microscopic atom-level relationship between H bubble formation and H trapping in vacancies is far from understood, i.e., it is still unclear that how the V-H$_{n}$ complexes grow to form H bubbles.

Particularly, the minimum energy of D ions for producing displacement damage in W is calculated to be 940 eV on the basis of the displacement threshold energy of 40 eV \cite{Tokunaga}, whereas the experimental results have shown that D plasmas with energy of tens of eV definitely produce blisters \cite{Miyamoto, Lindig, Shu1, Shu2, Shu3}. In sharp contrast with the case of W, the experimental result \cite{Myers} reported no bubbles formation in Pd implanted by 10 keV D to a very high supersaturation of about 1.7 D/Pd mole ratio. Shu thought that the lowered vacancy formation energy by trapping H might be responsible for the bubble formation in W when the incident H ion energy is greatly lower than the threshold value for the displacement \cite{Shu3}. Ogorodnikova \emph{et al.} proposed that several D atoms in a single vacancy could cause the displacement of neighboring lattice atoms due to stress-induced atomic diffusion, creating a divacancy and thus initiate bubble growth \cite{Ogorodnikova}. It is also found experimentally that the extent of blistering in W depends on the crystal orientation \cite{Miyamoto, Lindig, Shu1, Shu2, Shu3}. Similar to H, He irradiation also often leads to blister formation and subsequent degradation of the mechanical properties of metals \cite{Becquart}. Experimental results showed that bubbles are formed in W so far as the incident He ion energy is above 5 eV, which is the surface barrier potential energy for He penetrating into the W and is much lower than the threshold energy (500 eV) \cite{Nishijima}. Empirical potential studies in $\alpha$-Fe and Ni suggested that a spontaneous emission of a self-interstitial atom nearby He$_{n}$, V-He$_{n}$ and V$_{2}$-He$_{n}$ complexes is possible if $n$ is large enough \cite{Gao, Wilson}. Using first-principles methods, Fu \emph{et al}. investigated the energetics of V$_{m}$-He$_{n}$ complex in $\alpha$-Fe and predicted that the emission of a self-interstitial atom close to He$_{n}$ complex is energetically favorable for n$>$4 \cite{Fu1, Fu2}. Meanwhile, Henriksson \emph{et al.} suggested a mechanism for the growth of small He bubbles in low-energy He implanted W by performing molecular dynamics simulations: displacements of W atom nearby He$_{n}$ complex towards the surface via the formation of (111) crowdion interstitials \cite{Henriksson1, Henriksson2}.

Therefore, two crucial questions arise. The first question is how the energetics of nearest neighboring metal atoms of the V-H$_{n}$ and V-He$_{n}$ complexes change with the increase of $n$, in other words, what will happen to the neighboring metal atoms of the vacancy as the number of trapped H or He atoms increases? Does the creation of new vacancy become more easily at the neighboring sites closest to the trapped H/He vacancy with the number of H or He atoms? The second question is what are the similarity and difference in the local structures and energetics between V-H$_{n}$ and V-He$_{n}$ complexes. Here, the similarity should be related to the common feature that both H and He bubbles can be formed under low energy ions irradiation; the difference should be related to the different behaviors of H and He bubbles. The answers to the first question should play a dominant role in atomic-level understanding the H and He bubble formation mechanism; the answers to the second question will help us understand the difference (for example, in deposition depth) between H and He bubbles. In this paper, we have performed systematic first-principles calculations mainly to examine the energetics of relevant V-H/He$_{n}$ complexes in bcc W in hope of shedding light on the formation mechanism of H and He bubbles. In addition, some calculations in Pd (its stable H-site is octahedral interstitial site) have been carried out to make a comparison and explore the obvious difference mentioned above in H behaviors between W and Pd.
\section{Computation method}
The present calculations are performed within density functional theory as implemented in the VASP code with the projector augmented wave potential method \cite{Kresse}. The generalized gradient approximation and the Perdew-Wang functional are used to describe the electronic exchange and correlation effect \cite{Perdew}. The supercell composed of 128 lattice points ($4\times4\times4$) is used. The relaxations of atomic position and optimizations of the shape and size of the supecell are performed. The plane wave cutoff and k-point density, obtained using the Monkhorst-Pack method \cite{Monkhorst}, are both checked for convergence for each system to be within 0.001 eV per atom. Following a series of test calculations a plane wave cutoff of 500 eV is used and a k-point grid density of $3\times3\times3$ is employed. The structural optimization is truncated when the forces converge to less than 0.1 eV/{nm}. The vacancy, H(He) interstitial and substitutional defect formation energies are computed by following formula:
\begin{equation}
E_{f}=E_{tot}^{nW,mF}-nE^{W}-mE^{F},
\end{equation}
where $F$ indicates foreign H or He, $E_{tot}^{nW,mF}$ is the total energy of the system with $\emph{n}$ W atoms and $\emph{m}$ foreign atoms like H or He, $E^{W}$ is the energy per atom of pure crystal W and $E^{F}$ is one half of the energy of H$_{2}$ molecule (-3.40 eV) or the energy of an isolated He atom (0.00 eV). The binding energies of interstitial H or He atoms are determined for different configurations, which is expressed by:
\begin{equation}
 E_{b}^{F1,F2,\cdots, Fn}=\sum_{1}^{n}E_{tot}^{Fn}-E_{tot}^{F1+F2+\cdots+Fn}-(n-1)E_{tot}^{pure},
\end{equation}
where $E_{tot}^{Fn}$ is the energy of the W system with foreign atom $Fn$, $E_{tot}^{F1+F2+\cdots+Fn}$ is the energy of the system with foreign atoms from $F1$ to $Fn$, and $E_{tot}^{pure}$ is the total energy of pure crystal W. In such a scheme a positive binding energy indicates attractive interaction while a negative value means a repulsion. The trapping energy $E_{tr}^{V-F_{n}}$, when the number of H(He) atoms is increased from $n-1$ to $n$ in a vacancy, is defined as:
\begin{equation}
E_{tr}^{V-F_{n}}=E_{tot}^{V-F_{n}}-E_{tot}^{V-F_{n-1}}-(E_{tot}^{F_{tet}}-E_{tot}^{pure}),
\end{equation}
where \emph{n} is the number of $F$ atoms and $E_{tot}^{V-F_{n}}$ is the total energy of the system with $n$ $F$ atoms in a vacancy, $E_{tot}^{F_{tet}}$ is the total energy of the W system with a H(He) tetrahedral interstitial defect. A negative value of $E_{tr}^{V-F_{n}}$ indicates taking an interstitial H(He) atom and adding it to a vacancy that already contains $n-1$ H(He) atoms is energetically favorable, with $|E_{tr}^{V-F_{n}}|$ being the energy gained in that process. Here, we specially calculate the new vacancy formation energy of the W atom close to the V$_{m-1}$-F$_{n}$ complex using the following equation:
\begin{equation}
E_{f}^{V_{new}}=E_{tot}^{V_{m}-F_{n}}+E^{W}-E_{tot}^{V_{m-1}-F_{n}},
\end{equation}
where $E_{tot}^{V_{m}-F_{n}}$ is the total energy of the system with $m$ vacancies holding $n$ F atoms.
Zero point energy corrections are not taken into account, as it has very little influence (10$^{-2}$ eV) on our results such as binding energy, trapping energy and vacancy formation energy.

\section{Results and discussion}

\subsection{Binding properties of interstitial H-H and He-He pairs}

Our calculated defect formation energy results of H and He in the perfect W system are in good agreement with that previously reported \cite{Lee,Becquart2}: both H and He prefer to occupy tetrahedral interstitial site (TIS) rather than octahedral interstitial site (OIS), and the energy difference of H(He) at OIS and TIS is 0.39 eV(0.21 eV). So in the present work the interactions of two H(He) atoms located at different TIS separated by a certain distance are considered. The calculated binding energies as a function of the final distances of the two H(He) atoms in the W system are shown in Fig. 1. The results show that the binding energy increases with the increasing distance between the two H atoms, and fluctuates around zero when the distance is larger than 0.2 {nm} which agrees well with the results reported by Liu \emph{et al} \cite{Liu2}. The negative value of binding energy indicates the existence of repulsive interactions between near interstitial H atoms. The minimum binding energy is -0.46 eV in the W system, corresponding to the nearest distance of the two H atoms (0.16 {nm}). Due to the repulsive interactions of H atoms in the W system, H atoms can not form cluster easily but diffuse deeper into the bulk from the H-implanted W surfaces. In contrast, the binding energy is positive and decreases as the increase of the distance between the two He atoms, and fluctuates around zero when the distance is larger than 0.30 {nm}. The maximum binding energy is 1.08 eV when the two He atoms are separated apart by 0.15 {nm}, which is in good agreement with the results reported by Becquart \emph{et al} \cite{Becquart}. Moreover, it is noticeable that when the distance between the two He atoms ranges from 0.16 to 0.30 {nm}, they will aggregate together spontaneously to the distance of $\sim$0.15 {nm}. Therefore, He atoms can form cluster easily, which hinders the diffusion of He into the deep bulk.

\subsection{The effect of interstitial H/He atoms on the vacancy formation}

According to the relationship of defect concentration with temperature and defect formation energy \cite{Ismer}, at 300 K the equilibrium concentration of vacancy is relatively low ($\sim$10$^{-54}$) due to the large vacancy formation energy of 3.20 eV in the perfect W system. As pointed out above, under low-energy D or He ion irradiation (below threshold energy) no W atom is displaced to form a vacancy, however bubbles are observed at the W surface \cite{Miyamoto, Lindig, Shu1, Shu2, Shu3, Nishijima}. It is natural to firstly ask whether interstitial H and He atoms could result in substantial change in the vacancy formation energy. So, the effects of tetrahedral interstitial H and He on the vacancy formation have been studied. As shown in Table 1, the vacancy formation energies of the W atoms surrounded by 1, 2, 3 and 4 nearest H atoms at TIS are 1.99 eV, 0.83 eV, -0.35 eV and -1.39 eV, respectively. Meanwhile, the calculated binding energies suggest that the occupancy of two H atoms (the distance between the two H atoms is optimized to more than 0.2 {nm}) around the same W atom is possible, however the occupancy of 3 and 4 H atoms around the same W atom is difficult because of their repulsive interactions. In contrast, even one interstitial He atom can reduce the vacancy formation energy to -1.36 eV, suggesting the nearest W atom of interstitial He becomes unstable. The large positive binding energy indicates that there exist strong attractive interaction among the He atoms around the same W atom. Thus, both H and He atoms at TIS do reduce the energy required for the nearby vacancy formation considerably, especially He. Once the new vacancy is formed, it will change to the V-H$_{n}$ or V-He$_{n}$ complex.

\subsection{H/He-monovacancy interactions}

H and He diffuse with the barriers of as small as 0.2 eV \cite{Zhou} and 0.06 eV \cite{Becquart,Zhou} in perfect W, respectively, indicating that H and He can migrate quickly until they are tightly trapped by the defects to form H(He)-defect complex. In this part, our main objective is to explore the energetics related to the V-H$_{n}$(V-He$_{n}$) complex. We firstly calculate the trapping energy per H and He atom displayed in Fig. 2 as a function of the number of H and He trapped \emph{sequentially} in a single vacancy. For the case of H in single vacancy, with increasing number of H atoms the trapping energy shows a generally increasing trend, and its occasional fluctuations originate from the presence of H configurations with high symmetry. The atomic configurations of one to twelve H in the vacancy are in good agreement with the results reported by Ohsawa \cite{Ohsawa}. The side length of the unit cell of having trapped 12 H atoms inside the vacancy expands by $\sim$5\% compared to the perfect unit cell. And the formation of a H$_{2}$ molecule inside a vacancy is not observed. The Bader's charge analysis \cite{Bader} proves that H gains charge from surrounding W, and the averaged charge around H changes from $\sim$-0.54 $\mid$$e$$\mid$ to $\sim$-0.62 $\mid$$e$$\mid$, which results in the repulsive interactions among the H atoms inside the vacancy, and that is why H$_{2}$ molecule is not observed. It is energetically favorable for a W monovacancy to trap as many as 12 H atoms. The further calculation has been carried out of the binding energy between an additional H atom and the V-H$_{12}$ complex, which is shown in the inset of Fig. 2. The calculated binding energy clearly indicates that the presence of monotonically increasingly repulsive interaction with decrease of the distance between additional H atom and the V-H$_{12}$ complex. And when their distance is increased to at least 0.44 {nm} the binding energy approaches zero, suggesting that H atoms can not aggregate around the V-H$_{12}$ complex to grow and form H bubbles only based on the trapping H role of vacancy.

In contrast, the trapping energy of He in single vacancy is more negative than that of H in vacancy, indicating that He atoms are more strongly trapped in W vacancy than H atoms, which is consistent with the previously reported results that the binding of He and the vacancy is much stronger than that of H and the vacancy \cite{Zhou}. As displayed in Fig. 2, the trapping energy of He firstly increases rapidly from about -5 eV to about -3 eV and then fluctuates around -2.7 eV, being always far below zero energy even if He atoms is added up to 16. That is, He is extremely more favorable to aggregate in the vacancy rather than sit at the TIS far away from vacancy. Why can the vacancy trap so many He atoms? The optimized structure configurations from one to sixteen He atoms trapped in the vacancy are shown \emph{on the same scale} in Fig. 3. Obviously, with increasing He atoms the systems expand and distort more and more strongly but the nearest distance of He-He keeps about 0.16 nm. And 15th and 16th He atoms indeed move out of the original unit cell with the vacancy, indicating that the vacancy can trap up to 14 He atoms and additional He atoms prefer to cluster round the V-He$_{14}$ complex, which is in agreement with the tendency to form He clusters confirmed in Ref. \cite{Becquart}. Here it should be stressed that the binding energy of V-He$_{14}$ complex with additional He atom is more than one eV larger than the strongest binding energy between two interstitial He atoms. Therefore He atoms may aggregate persistently inside/around vacancy to grow and form He bubbles. The unit cell of having trapped 14 He atoms expands by $\sim$26\% in length compared to the perfect unit cell, indicating that the swelling from He atoms is very heavy. There exist high symmetry configurations for the cases of 1, 2, 3, 4, 6 and 8 He atoms, which are partially responsible for the fluctuation in the trapping energy with the number of He atoms.

Based on the above obtained results: the existence of respective repulsion- and attraction-interactions of interstitial H pairs and interstitial He pairs in bulk W, single interstitial He atom yielding the negative vacancy formation energy while single interstitial H atom leading to the reduced but still positive vacancy formation energy, and the appearance of repulsive interaction between additional H and V-H$_{12}$ complex and strongly attractive interaction of additional He with V-He$_{14}$ complex, we may draw the following conclusions. During He atom diffuses into the bulk it can be easily attracted to the V-He$_n$ complex or the other He atoms. Whereas during H atoms diffuse into the bulk, because of the absence of the attractive force from the V-H$_{12}$ complex or the other H atoms, H atoms can diffuse deeper into the bulk than He atoms. Thus it could be understandable that even at temperatures where the migration rate of He is far larger than that of H at 500 K, He will form bubbles right at $\sim$10{nm} to the W surface, while H bubbles are found at micrometer depths \cite{Nicholson, Sze1, Haasz, Sze2, Iwakiri, Alimov, Henriksson3}.

\subsection{Mutation from V-H/He$_{n}$ complex into V$_{2}$-H/He$_{n}$ to lead to the growth of H/He-vacancy complex}

Although the above obtained results clearly reveal that a single vacancy in W can trap as many as 12 H or 14 He atoms, it remains unclear how the V-H$_{n}$ and V-He$_{n}$ complexes grow to form H and He bubbles specially due to the saturation of H trapped inside vacancy. Using Eq. (4), we systematically calculate the new (second) neighboring vacancy formation energy of the W atom closest to the vacancy trapped $n$ H or He atoms (i.e., V-H$_{n}$ or V-He$_{n}$ complex). Note that the first nearest neighbor ($1nn$), second nearest neighbor ($2nn$) and third nearest neighbor ($3nn$) (see the inset of Fig. 4) vacancy formation energies around the already existed vacancy are 3.16 eV, 3.52 eV and 3.22 eV, respectively. Thus, it is much difficult that the vacancy grows to form large vacancy clusters spontaneously under low energy ions irradiation. However, to our surprise, after the already existed vacancy having trapped H or He atoms the situation will be very different. As shown in Fig. 4, the $1nn$ and $2nn$ vacancy formation energies are displayed as a function of the number of trapped H or He atoms inside the vacancy. The $1nn$ and $2nn$ vacancy formation energies of the V-H$_{n}$ complex reduce in a steplike way, slowly at the first (i.e., when the number of H is between 1-5) and then decrease very rapidly to $\sim$0 eV when the trapped H number increases from 6 to 9. It should be stressed that the substantial reduction in $1nn$ and $2nn$ vacancy formation energies presently observed is quite different from the previously reported vacancy formation energies in metals due to the insertion of H \cite{Fukai4, Zhang}, where the energy of a vacancy is lowered mainly by the sum of binding energies of H atoms with vacancy. In general, the $1nn$ vacancy formation energy is smaller than the $2nn$ vacancy formation energy. Compared to the remarkably change in the $1nn$ and $2nn$ vacancy formation energies, the $3nn$ vacancy formation energy decreases very weakly, here which is not presented in Fig. 4 for clarity.

In sharp contrast, as shown in Fig. 5 we have not observed much strong decrease in the $1nn$ and $2nn$ new vacancy formation energies of H-vacancy complex in Pd. It is found that the maximum of 6 H atoms can be held in a vacancy in Pd and the configurations of the H atoms are in good agreement with previous results \cite{Vekilova}. The resulting neighboring vacancy formation energy due to the trapped 6 H atoms is still larger than 1.2 eV, therefore, it is reasonable that as pointed out previously no bubbles form in Pd implanted by 10 keV D ions \cite{Myers}.

For the case of V-He$_{n}$ complex as shown in Fig. 4, the $1nn$ and $2nn$ vacancy formation energies decrease sharply and almost linearly before He inside vacancy adds up to 10 and then do not change obviously. It should be pointed out that both $1nn$ and $2nn$ vacancy formation energies are lower than 0 eV when the number of trapped He is beyond 4. The $3nn$ vacancy formation energy closest to the V-He$_{n}$ complex behaves like that of the V-H$_{n}$ complex, being unsensitive to the trapped He atom number. The great difference in energetics between H and He trapped in a single vacancy is in that the $1nn$ and $2nn$ vacancy formation energies decease to $\sim$0 eV when the trapped atom number is up to 9 for H whereas the trapped number is larger than 4 for He. The reasons for the substantial reduction of new vacancy formation energy near the V-H$_{n}$ and V-He$_{n}$ complexes can be understood by the weakened W-W metal bond which originates from two parts: the increased W-W bond length and the decrease of electron density between W-W atoms around the corresponding complex. The averaged nearest-neighbor distances of both the $1nn$ and $2nn$ W atoms increase by $\sim$0.005 {nm} for the V-H$_{n}$ complex and $\sim$0.012 {nm} for the V-He$_{n}$ complex when $n$ changes from 1 to 12. The increase of the W-W bond length weakens the W-W interactions, causing the reduction of vacancy formation energies nearby V-H$_{n}$ and V-He$_{n}$ complexes. The electron density around $1nn$ and $2nn$ W atoms is obviously reduced. The electron densities of (110) plane across $1nn$, $2nn$ and $3nn$ atoms (named by 1, 2 and 3 in Fig. 4) of the vacancy are calculated and displayed in Fig. 6. Specially, we take two different cases of the V-H$_{10}$ and V-He$_{6}$ complexes for example, and they are compared with the `empty' vacancy (Fig. 6(a)). As shown in Fig. 6(b), the accommodation of 10 H atoms in the single vacancy directly results in the extension of the light blue region, and the shrinking of the dark green region around the $1nn$ and $2nn$ W atoms, while the various color regions around the $3nn$ W atoms do not show evident changes. These indicate that the electron density around $1nn$ and $2nn$ W atoms reduces obviously, whereas the electron density around the $3nn$ W atoms shows little change. Similar phenomena are found for the V-He$_{6}$ complex that the electron density round the $1nn$ and $2nn$ W atoms decreases obviously (shown in Fig. 6(c)), and some dark green regions even disappear, but the electron density around the $3nn$ W atoms of the V-He$_{6}$ complex changes little. The reduction of the electron density could further weaken the interactions of W-W, leading to the decrease of the $1nn$ and $2nn$ vacancy formation energies nearby both V-H$_{10}$ and V-He$_{6}$ complexes.

From above results and discussion, we can conclude that the new vacancy is much easily produced in the region closest to the V-H$_{n}$ and V-He$_{n}$ complexes when the number of H or He inside the vacancy is beyond a certain number. This means that the V-H$_{n}$ and V-He$_{n}$ complexes can easily (even spontaneously) mutate into the V$_{2}$-H$_{n}$ and V$_{2}$-He$_{n}$ complexes when $n$ is large enough, respectively. The further calculations have been performed about the energetics of closest W atoms (nearby V$_{2}$-H$_{10}$ (V$_{2}$-He$_{10}$) complex) which are removed in a stepwise fashion to create V$_{3}$-H$_{10}$(V$_{3}$-He$_{10}$) and V$_{4}$-H$_{10}$(V$_{4}$-He$_{10}$) complexes, we find that the successive vacancy formation energies to form these complexes are 1.87 eV(-2.26 eV) and 2.42 eV(-0.43 eV), respectively. If the trapped H or He atoms are larger than 10, these new vacancy formation energies will be further reduced. This finding suggests a cascade mechanism, as recently reported in the large variation of vacancy formation energies in the surface of crystalline ice \cite{Watkins}, whereby once a vacancy is created and when this vacancy traps certain numbers of H or He atoms, neighboring W atoms become very weakly bound and thus easily to be removed to form a new vacancy, and with the potential to lead to the growth of H/He-vacancy complexes.

A V-He-complex mutation growth mechanism for He bubble has been mentioned by Caspers et al. in 1978 \cite{Caspers}, which works as follows. Assuming the He atoms are trapped in a single vacancy and form V-He$_{n}$ complex due to the strong He-vacancy bonding energy. Some fraction of the V-He$_{n}$ complexes, which reach a critical size, mutate into a complex with two vacancies (V$_{2}$-He$_{n}$) by ejecting an interstitial into the metal matrix. By absorbing He atoms and further ejecting interstitial metal atoms, the complexes become larger and larger and finally lead to He bubble formation. Our results indicate that the mechanism is also suitable for H. The presently observed substantial reduction of the $1nn$ and $2nn$ vacancy formation energies close to the V-H$_{n}$ and V-He$_{n}$ complexes, to our best knowledge, not only gives the direct evidence for this mechanism, but also gives the reasonable explanation of the experimental results: Why H and He bubbles with diameters of a few to hundreds of microns could form on W surface even if the ion energy is so low that no displacement damage is created \cite{Miyamoto, Shu2, Sze1, Haasz, Sze2, Iwakiri}.

\section{Conclusions}
In summary, based on the first-principles method we have investigated the energetics of H/He-vacancy complex in W by calculating the trapping energy of H/He and the new nearby vacancy formation energy. We find that a monovacancy can accommodate up to 12 H and 14 He to form V-H$_{12}$ and V-He$_{14}$ complexes, respectively. And the V-H$_{12}$ exhibits strong repulsive role with the approach of additional H atoms, but the V-He$_{14}$ shows great attraction to the nearby He atoms. The aggregation of H and He in W vacancy remarkably favors the creation of new vacancy around the H/He-vacancy complexes: the first-nearest-neighbor and second-nearest-neighbor formation energies of vacancy close to the H/He-vacancy complex decrease to about 0 eV when the trapped atom number is up to 9 for H and larger than 4 for He. These results, not only provide the direct evidence of the He-vacancy complex mutation mechanism proposed by Caspers for the He bubble formation, but also suggest a cascade mechanism, as recently reported in the large variation of vacancy formation energies in the surface of crystalline ice by Watkins et al, whereby once a vacancy is created and when this vacancy traps certain numbers of H or He atoms, neighboring W atoms become very weakly bound and thus easily to be removed to form a new vacancy, and with the potential to lead to the growth of H/He-vacancy complexes. Besides, the results well explain the experimental phenomena --- the huge discrepancy of deposition depth of H and He in W, and the formation of H/He bubble with diameters of a few to hundreds of microns on W surface even if the ion energy is so low that no displacement damage is created. However, there is no quite large decrease in the new neighboring vacancy formation energy nearby a vacancy having trapped H atoms in Pd, leading to the neighboring vacancy formation energy still being larger than 1.2 eV, thus no bubbles formation in Pd even implanted by 10 keV D ions.

\section*{Acknowledgement}

This work was supported by the National Magnetic Confinement Fusion Program (Grant Nos.: 2011GB108004 and 2009GB106005), the National Natural Science Foundation of China (Nos.: 91026002, 91126002) and the Strategic Priority Research Program of Chinese Academy of Sciences
(Grant Nos.: KJCX2-YW-N35 and XDA03010303), and by the Center for Computation Science, Hefei Institutes of Physical Sciences.






\newpage
\textbf{Figure caption: }

Fig. 1. The binding energy as a function of the final distance of interstitial H-H(He-He) pair in the perfect W system. The diatomic He will aggregate together to a distance of $\sim$0.15 {nm} when the two He atoms are separated by from 0.16 to 0.30 {nm}. The inset is the enlargement of the binding energy of two He atoms separated by from 0.143 to 0.154 {nm}.

Fig. 2. Trapping energy per H and He in a single W vacancy as a function of the number of trapped H and He atoms, here the zero point is the energy of H or He at the TIS far away from the vacancy. Inset: the binding energy between additional H atom with the V-H$_{12}$ complex as a function of the distance between additional H atom and the center of the V-H$_{12}$ complex.

Fig. 3. The lowest-energy configurations of 1 to 16 He atoms inside (or around) a single W vacancy, note that here all configurations are shown using the same scale. Big and small balls indicate W and He atoms, respectively.

Fig. 4. The $1nn$ and $2nn$ vacancy formation energy of the V-H$_{n}$ and V-He$_{n}$ complexes as a function of the number of trapped H or He atoms. Lines are guides to the eyes.

Fig. 5. The $1nn$ and $2nn$ vacancy formation energy of the V-H$_{n}$ complex as a function of the number of trapped H atoms in Pd. Lines are guides to the eyes. Inset: trapping energy per H in a single Pd vacancy as a function of the number of H atoms, here the zero point is the energy of H at the OIS far away from the vacancy. From the inset, a maximum of 6 H atoms could be held in a vacancy in Pd.

Fig. 6. The electron density maps (electron/{\AA}$^{3}$) of three different cases: the empty vacancy (a), the V-H$_{10}$ (b) and V-He$_{6}$ (c) complexes. In all cases, the slices are cut through the same (110) plane of the supercell considered. 1, 2 and 3 denote the first, second and third nearest neighbor W atoms of the vacancy. `x' and `y' represent the directions of  $\langle$\={1}01$\rangle$ and $\langle$0\={1}0$\rangle$.

\newpage
Table 1 The vacancy formation energy $E_{f}^{V}$ (eV) of W atom that has multiple (1-4) neighboring interstitial H(He) atoms are calculated. Meanwhile, the binding energy $E_{b}$(eV) of these H(He) atoms are also calculated using Eq. 2.
\begin{center}
\begin{tabular}{ccccccccccc}
\hline
$$          &1H & 2H & 3H  & 4H &  1He & 2He & 3He  & 4He \\
\hline
$E_{b}$(eV)       & --     & 0.00   & -0.06  & -0.22  & --      & 1.08   & 2.18   & 3.61 \\
$E_{f}^{V}$(eV)   &1.99    & 0.83   & -0.35  & -1.39  &-1.36    & -3.38  & -5.31  & -6.64\\
\hline
\end{tabular}
\end{center}

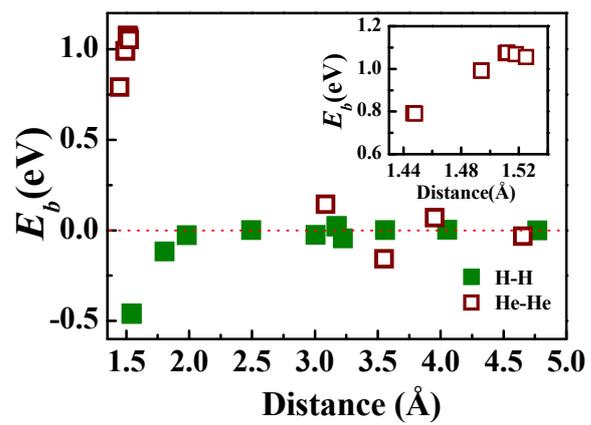
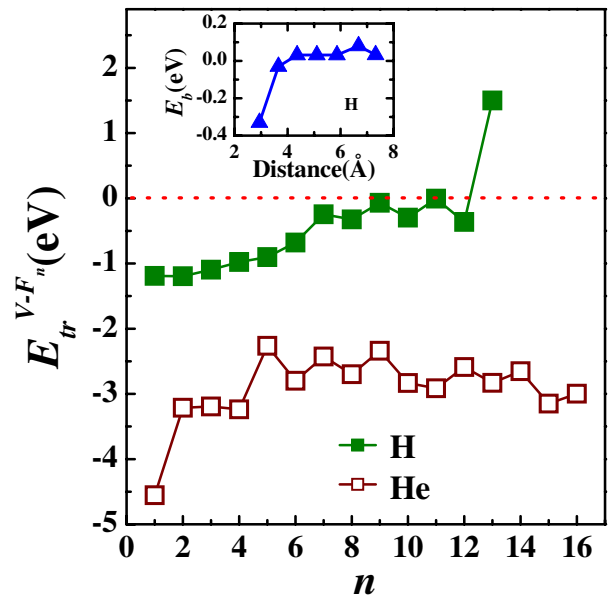
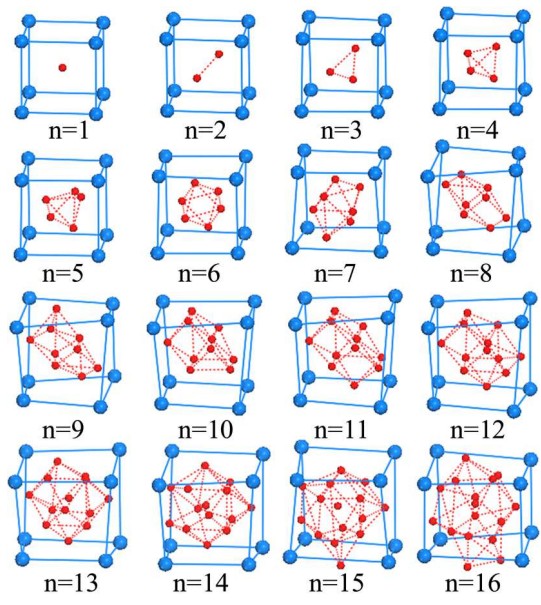
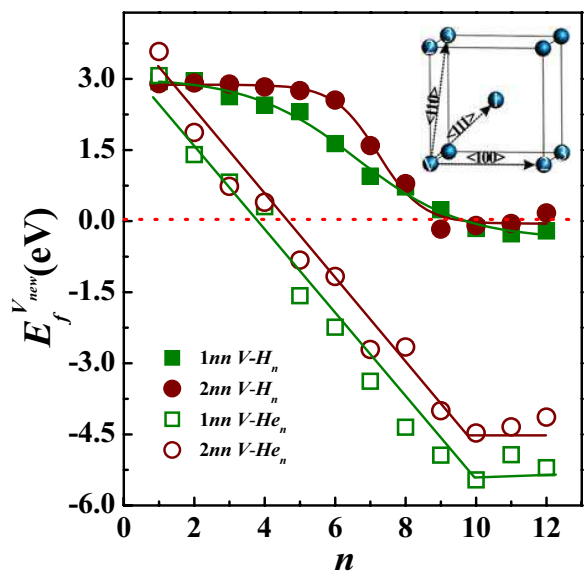
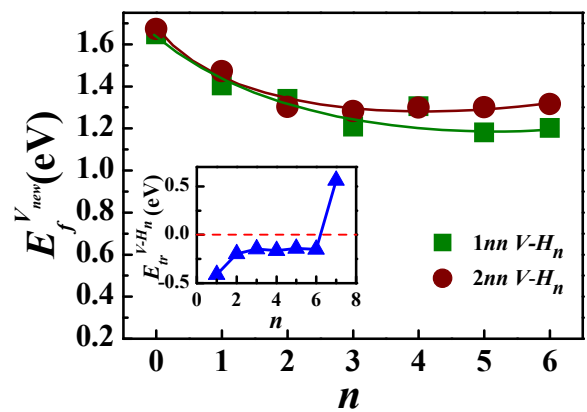
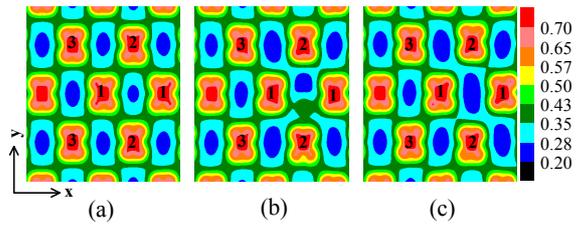
\end{document}